\documentclass[aps,showpacs,11pt]{revtex4}
\usepackage{dcolumn}
\usepackage{graphicx}
\usepackage{amsmath}
\usepackage{amsfonts}
\usepackage{amssymb}
\usepackage{psfrag}
\usepackage{wrapfig}
\usepackage{subfigure}
\usepackage{makeidx}
\usepackage{bm}
\usepackage{epsf}
\usepackage{hyperref}
\usepackage{color}
\usepackage{multirow,dcolumn}
\usepackage{graphicx}

\begin{document}

\title{Subregion complexity and confinement-deconfinement transition in a holographic QCD model}

\author{Shao-Jun Zhang}
\email{sjzhang84@hotmail.com}
\affiliation{Institute for Advanced Physics and Mathematics, Zhejiang University of Techonology, Hangzhou 310023, China}

\affiliation{Center for Gravitation and Cosmology, College of Physical Science and Technology,
Yangzhou University, Yangzhou 225009, China}
\date{\today}

\begin{abstract}

We study the subregion complexity in a semi-analytical holographic QCD model. Two cases with different warped factor are considered and both can realize confinement-deconfinement transition. By studying the behavior of the renormalized holographic complexity density $\hat{\cal C}$ versus the subregion length scale $\ell$, we find that for both cases, $\hat{\cal C}$ always experiences a discontinuity at certain critical value $\ell_c$ in confinement phases, while it is always continuous in deconfinement phases. This property may be seen as a signal to characterize confinement or deconfinement phases. The behavior of $\hat{\cal C}$ versus the temperature and chemical potential is also investigated and our results show that $\hat{\cal C}$ exhibits behavior characterizing the type of the transition. That is, it experiences a discontinuity at the transition temperature for $\mu < \mu_c$ where first-order confinement-deconfinement phase transition happens, while it is always continuous for $\mu> \mu_c$ where the transition turns into a turnover. These results imply that the renormalized holographic complexity density may be used as a good parameter to characterize the corresponding phase structures.

\end{abstract}


\maketitle

\section{Introduction}

Past decades have witnessed the great successes of application of AdS/CFT~\cite{Maldacena:1997re,Gubser:1998bc,Witten:1998qj} or the more generic gauge/gravity duality in various areas of modern theoretical physics, such as condensed matter physics (CMT)~\cite{Hartnoll:2009sz,Herzog:2009xv,McGreevy:2009xe,Horowitz:2010gk,Cai:2015cya}, QCD~\cite{Mateos:2007ay,Gubser:2009md,CasalderreySolana:2011us}, quantum information theory (QIT)~\cite{Swingle:2009bg,Swingle:2012wq,Qi:2013caa} and cosmology~\cite{Banks:2004eb} and etc. By connecting physical quantities in the boundary quantum field theories (QFTs) to certain geometric quantities in the gravity side in the bulk, it may help deepening our understanding of both the strongly coupled problems in the QFTs side as well as the origin of spacetime in the gravity side. To achieve this goal, one should try to find the dictionary to know how the two sides are connected exactly. After decades' efforts of physicists, several concrete correspondences between the two sides are proposed. The most recent one is trying to connect the complexity in the boundary QFTs to some geometric quantity in the gravity side~\cite{Susskind:2014rva}. In QFTs (or QIT), complexity of a state is an important concept defined as the minimum number of unitary operators (or gates) needed to produce the state of interest from some reference state. At present, there are two concrete proposals, namely the CV (complexity=volume) and CA (complexity=action) conjectures. In the CV conjecture, complexity of the boundary state living on a time slice $\Sigma$ equals to the extremal volume of a codimension-one hypersurface ${\cal B}$ in the bulk ending on $\Sigma$ at the boundary~\cite{Stanford:2014jda}, that is
\begin{eqnarray}\label{CV}
C_V (\Sigma) =\bm{\max}_{\partial {\cal B} = \Sigma} \left(\frac{{\rm Vol} ({\cal B})}{G_{d+1} R}\right),
\end{eqnarray}
where $G_{d+1}$ is the gravitational constant in $(d+1)$-dimensional asymptotical AdS spacetime and $R$ is some length scale characterizing the bulk geometry, for example the AdS radius or the horizon radius. The CA conjecture states that complexity of the boundary state equals to the on-shell gravitational action on the so-called Wheeler-DeWitt (WDW) patch of the bulk spacetime~\cite{Brown:2015bva,Brown:2015lvg}. Each conjecture has its own advantages and disadvantages~\cite{Hashimoto:2018bmb}. Inspired by these ideas, there raises an intensive interest in studying the holographic complexity for various holographic gravity models to check these proposals~\cite{Momeni:2016ekm,Cai:2016xho,Brown:2016wib,Couch:2016exn,Yang:2016awy,Chapman:2016hwi,
Carmi:2016wjl,Pan:2016ecg,Brown:2017jil,Kim:2017lrw,Cai:2017sjv,Alishahiha:2017hwg,Bakhshaei:2017qud,
Tao:2017fsy,Guo:2017rul,Zangeneh:2017tub,Alishahiha:2017cuk,Abad:2017cgl,Reynolds:2017lwq,Hashimoto:2017fga,Nagasaki:2017kqe,Miao:2017quj,Ge:2017rak,
Ghodrati:2017roz,Qaemmaqami:2017lzs,Carmi:2017jqz,Kim:2017qrq,Cottrell:2017ayj,Sebastiani:2017rxr,
Moosa:2017yvt,HosseiniMansoori:2017tsm,Reynolds:2017jfs,Chapman:2018dem,Chapman:2018lsv,Khan:2018rzm,Caputa:2018kdj}.

It should be noted that the above two conjectures of holographic complexity are proposed for the whole boundary system. They both have been extended to be defined on subsystem respectively in Refs.~\cite{Alishahiha:2015rta} and~\cite{Carmi:2016wjl}, and one may call them holographic subregion complexity. In the subregion CV proposal, complexity of a subregion $B$ of the boundary system equals to the volume of the extremal codimension-one hypersurface $\Gamma_B$ enclosed by $B$ and the corresponding Hubney-Ryu-Takayanagi (HRT) surface $\gamma_B$~\cite{Ryu:2006bv,Hubeny:2007xt}, that is
\begin{eqnarray}\label{subCV}
{\cal C} (B) = \frac{{\rm Vol (\Gamma_B)}}{8 \pi G_{d+1} L}.
\end{eqnarray}
Here $L$ is the AdS radius and the constant factor $8\pi$ is irrelevant but just a convention. Actually, it has been suggested to be dual to the fidelity susceptibility in QIT~\cite{Alishahiha:2015rta,MIyaji:2015mia}. While in the subregion CA proposal, complexity of subregion $B$ equals to the on-shell gravitational action on the intersection region between WDW patch and the so-called entanglement wedge~\cite{Czech:2012bh,Headrick:2014cta}. There are also lots of work and effort devoted to understand the holographic subregion complexity~\cite{Caputa2017,Caputa2017b,Czech1706,subBenAmi2016,subRoy2017,
subBanerjee2017,subBakhshaei2017,subSarkar2017,subZangeneh2017,subMomeni2017,subRoy2017b,subCarmi2017,Chen:2018mcc,Ageev:2018nye}.

Our main goal in this work is to investigate the holographic subregion complexity in a class of holographic models intending to mimic the real QCD physics. As we have already known for a long time, it is an extremely challenging problem to understand QCD physics at low energy regime where quarks are confined. As a strong/weak duality, AdS/CFT may provide us powerful tools to deal with this strongly-coupled problem by mapping it to a weakly-coupled and simpler gravitational problem, and thus may help us to understand essential properties of QCD, especially the confinement-deconfinement transition. There have been already several holographic QCD models, either from a top-down approach which can be reduced from a more fundamental theory (string theory for examaple) or from a bottom-up approach which is constructed phenomenologically to fit lattice results, including Gubser's model~\cite{Gubser:2008ny,Gubser:2008yx,DeWolfe:2010he}, the improved holographic QCD model (IHQCD)~\cite{Gursoy:2007cb,Gursoy:2007er,Gursoy:2008za,Kiritsis:2009hu,Gursoy:2009jd} and the semi-analytical holograpic QCD model~\cite{Li:2011hp,Cai:2012xh,He:2013qq,Yang:2014bqa,Yang:2015aia,Dudal:2017max,Dudal:2018ztm}. Each model can realize part key properties of the real QCD, but none can realise the whole QCD physics. In this work, we will focus on the semi-analytical holographic QCD model. It is an Einstein-Maxwell-dilation system having the merit that it allows semi-analytical black hole solutions which can be used to study the dual QCD physics at finite temperature. The black hole solutions are determined by two unfixed functions, one is the gauge kinetic function $f(\phi)$ representing the coupling between the gauge field and the dialton and the other is the warped factor $A(z)$ describing deformation from the standard AdS spacetime. By choosing appropriate forms of the two functions, this model can realise the vector meson spectrum holographically which agrees well with the lattice result. Moreover, it can also realise the confinement-deconfinement transition for at least two choices of the warped factor. In Ref.~\cite{Dudal:2018ztm}, the authors apply the holographic entanglement entropy (HEE) to probe the confinement-deconfinement transition in this model. Their results show that HEE may be a good parameter to characterize the corresponding phase structures. Also see Ref.~\cite{Zhang:2016rcm} for a related work but in Gubser's model. As there is a deep connection between entanglement entropy and complexity, it is natural and interesting to see if the complexity can also be used to probe this transition. In Ref.~\cite{Zhang:2017nth}, in Gubser's model, with the CA conjecture we show that holographic complexity can be a good parameter to characterize the corresponding phase structures (see also Ref.~\cite{Ghodrati:2018hss}). In this work, we will apply the subregion CV conjecture Eq.~(\ref{subCV}). As we will show in the main context, the holographic subregion complexity can be used to probe the confinement-deconfinement transition as well.

The work is organized as follows. In Sec. II, we will briefly review the semi-holographic QCD model. Then we discuss its thermodynamics and phase structures in Sec. III. In Sec. IV, we study the holographic subregion complexity with the subregion CV conjecture to probe the confinement-deconfinement transition. The last section is devoted to summary and discussions.

\section{Holographic QCD Model}

In this section, we briefly review the holographic QCD model that has been previously studied thoroughly in Refs.~\cite{Li:2011hp,Cai:2012xh,He:2013qq,Yang:2014bqa,Yang:2015aia,Dudal:2017max,Dudal:2018ztm}. In Einstein frame, the model is described by a five-dimensional Einstein-Maxwell-dilaton action as
\begin{eqnarray}
S = \frac{1}{16\pi G_5} \int d^5x \sqrt{-g} \left[R - \frac{f(\phi)}{4} F_{\mu\nu}F^{\mu\nu} -\frac{1}{2}\partial_\mu \phi \partial^\mu \phi-V(\phi)\right],
\end{eqnarray}
where $G_5$ is the Newton constant in five dimensions. The Maxwell field is coupled to the dilaton through the gauge kinetic function $f(\phi)$. $V(\phi)$ is the dilaton potential. From the action, the equations of motion can be derived
\begin{eqnarray}
R_{\mu\nu} -\frac{1}{2} g_{\mu\nu} R &=& \frac{f(\phi)}{2}\left(F_{\mu\rho}F_\nu^{~\rho} -\frac{1}{4}g_{\mu\nu}F^2\right)+\frac{1}{2}\left[\partial_\mu \phi \partial_\nu \phi -\frac{1}{2} g_{\mu\nu} (\partial\phi)^2 - g_{\mu\nu} V(\phi)\right],\\
\nabla_\mu[f(\phi) F^{\mu\nu}]&=&0,\\
\nabla^2\phi &=& \frac{\partial V}{\partial \phi} + \frac{F^2}{4}\frac{\partial f}{\partial \phi}.
\end{eqnarray}

To study QCD at finite temperature holographically, we need to find black hole solutions. To achieve this goal, we assume the following ansatz
\begin{eqnarray}
ds^2 &=& \frac{e^{2 A(z)}}{z^2} \left[-g(z) dt^2 + \frac{dz^2}{g(z)} + d\vec{x}^2 \right],\\
A_\mu &=& (A_t(z),0,\vec{0}),\quad \phi=\phi(z),
\end{eqnarray}
where the horizon $z_H$ is given by the smallest root of the equation $g(z_H)=0$, and $z=0$ corresponds to the conformal boundary. For the sake of simplicity, we have set the radius of $AdS_5$ to be unit by scale invariance.\\
With this ansatz, the equations of motion become
\begin{eqnarray}
A''+3 A'^2 + \left(\frac{3 g'}{2 g} - \frac{6}{z}\right) A' - \frac{1}{z} \left(\frac{3 g'}{2 g} - \frac{4}{z}\right) + \frac{g''}{6 g} + \frac{e^{2 A} V}{3 z^2 g} &=& 0,\\
A'' - A'^2 + \frac{2}{z} A' + \frac{\phi'^2}{6} &=& 0,\\
g'' +\left(3 A'-\frac{3}{z}\right) g' - e^{-2 A} z^2 f A_t'^2 &=& 0,\\
A_t'' + \left(\frac{f'}{f} + A' - \frac{1}{z}\right) A_t' &=& 0,\\
\phi'' + \left(\frac{g'}{g} + 3 A' - \frac{3}{z}\right) \phi' + \left(\frac{z^2 e^{-2 A} A_t'^2}{2 g} \frac{\partial f}{\partial \phi} - \frac{e^{2 A}}{z^2 g} \frac{\partial V}{\partial \phi}\right) &=& 0.
\end{eqnarray}
To solve the above equations of motion, specific boundary conditions are needed. We expect that the metric in the string frame to be asymptotic to $AdS_5$ at the conformal boundary and the black hole solutions are regular at the horizon. In Einstein frame, these conditions turn out to be
\begin{eqnarray}
&&A(0) + \sqrt{\frac{1}{6}} \phi(0) =0,\quad g(0) = 1,\\
&&A_t(z_H) = g(z_H) =0.
\end{eqnarray}
The temporal part of the gauge field $A_t$ has the following behavior at the conformal boundary
\begin{eqnarray}
A_t (z) = \mu +\rho z^2 + {\cal O}(z^4),
\end{eqnarray}
where, according to the AdS/CFT dictionary, $\mu$ and $\rho$ correspond to the chemical potential and baryon density of the dual QCD respectively.

To solve the equations of motion, we should first know the exact form of the gauge kinetic function $f(\phi)$. With the requirement of producing the linear vector meson spectrum, $f(\phi)$ can be chosen to be a simple form
\begin{eqnarray}
f(z) = e^{-c z^2 - A(z)},
\end{eqnarray}
with this choice of $f(\phi)$, the equations of motion can be solved analytically as
\begin{eqnarray}
g(z) &=& 1- \frac{1}{\int_0^{z_H} y^3 e^{-3 A} dy} \left[\int_0^z y^3 e^{-3 A} dy - \frac{2 c\mu^2}{\left(1-e^{c z_H^2}\right)^2}\left|
\begin{array}{ll}
\int_0^{z_H} y^3 e^{-3 A} dy & \int_0^{z_H} y^3 e^{-3 A} e^{c y^2} dy\\
\int_{z_H}^z y^3 e^{-3 A} dy & \int_{z_H}^z y^3 e^{-3 A} e^{c y^2} dy
\end{array}\right|\right],\nonumber\\
\\
A_t(z) &=& \mu \frac{e^{c z^2} - e^{c z_H^2}}{1 - e^{c z_H^2}},\\
\phi'(z) &=& \sqrt{-6 \left(A'' - A'^2 + \frac{2}{z} A'\right)},\\
V(z) &=& -3 z^2 g e^{-2 A} \left[A'' + 3 A'^2 + \left(\frac{3 g'}{2 g} - \frac{6}{z}\right) A' - \frac{1}{z} \left(\frac{3 g'}{2 g} - \frac{4}{z}\right) + \frac{g''}{6 g}\right].
\end{eqnarray}
From the metric, we can derive the Hawking temperature of the black hole
\begin{eqnarray}
T_{\rm H} = \frac{z_H^3 e^{-3 A(z_H)}}{4\pi \int_0^{z_H} y^3 e^{-3 A(y)} dy} \left[1 + \frac{2 c\mu^2 \left(e^{-c z_H^2} \int_0^{z_H} y^3 e^{-3 A(y)} dy - \int_0^{z_H} y^3 e^{-3 A(y)} e^{-c y^2} dy\right)}{\left(1 - e^{-c z_H^2}\right)^2}\right].
\end{eqnarray}
From the above expressions, we can see that the black hole solution is fixed once the form of the warped factor $A(z)$ is given. In the limit $z_H \rightarrow \infty$, $g(z)=1$ and thus the above black hole solutions reduce to a thermal-AdS solution. However, as emphasized in Refs.~\cite{Dudal:2017max,Dudal:2018ztm}, a little different from the standard thermal-AdS solution, this thermal-AdS solution has a non-trivial bulk structure because of having a non-zero warped factor.

\section{Thermodynamics and phase diagram}

In this section, we will discuss the thermodynamics of the black hole solutions obtained in the last section. From it, the phase diagram of the dual QCD can be reproduced. Before doing so, we should first fix the parameter $c$ and the exact form of the warped factor $A(z)$.

With this model, one can obtain the vector meson mass spectrum holographically~\cite{He:2013qq} which depends on the value of the parameter $c$. And by comparing to the experimental data on the mass of the lowest lying heavy meson states, the parameter $c$ can be fixed as
\begin{eqnarray}
c \simeq 1.16 {\rm GeV}^2.
\end{eqnarray}

For the exact form of the warped factor $A(z)$, we may have many choices. In the next two subsections, we will follow Refs.~\cite{Dudal:2017max,Dudal:2018ztm} and consider two specific choices.

\subsection{Case I: Standard confinement/deconfinement phases}

In this case, the warped factor $A(z)$ takes the following simple form
\begin{eqnarray}
A(z) = -\bar{a} z^2,
\end{eqnarray}
with $\bar{a} = c/8 \simeq 0.145$.

\begin{figure}[!htbp]
\centering
\subfigure[$~~{\rm Case~I}$]{\includegraphics[width=0.45\textwidth]{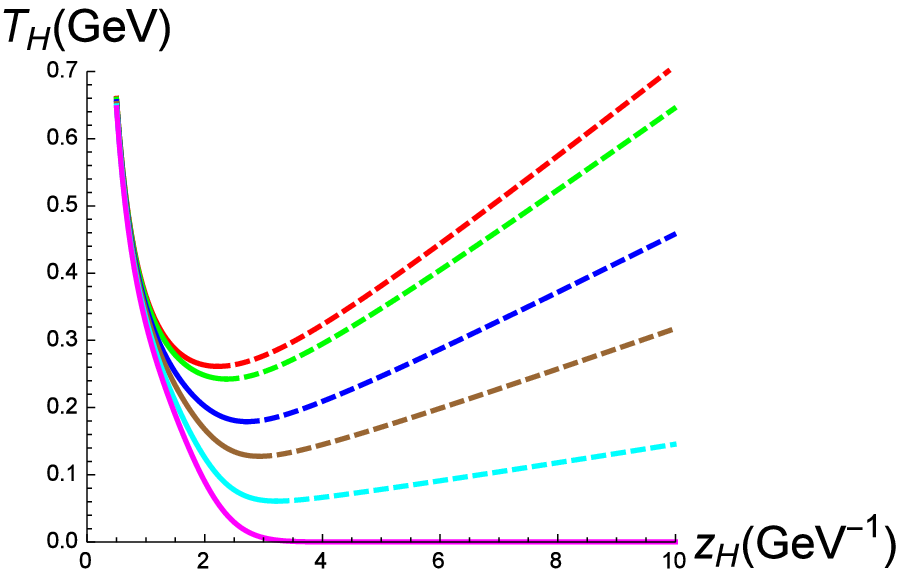}}\quad
\subfigure[$~~{\rm Case~II}$]{\includegraphics[width=0.45\textwidth]{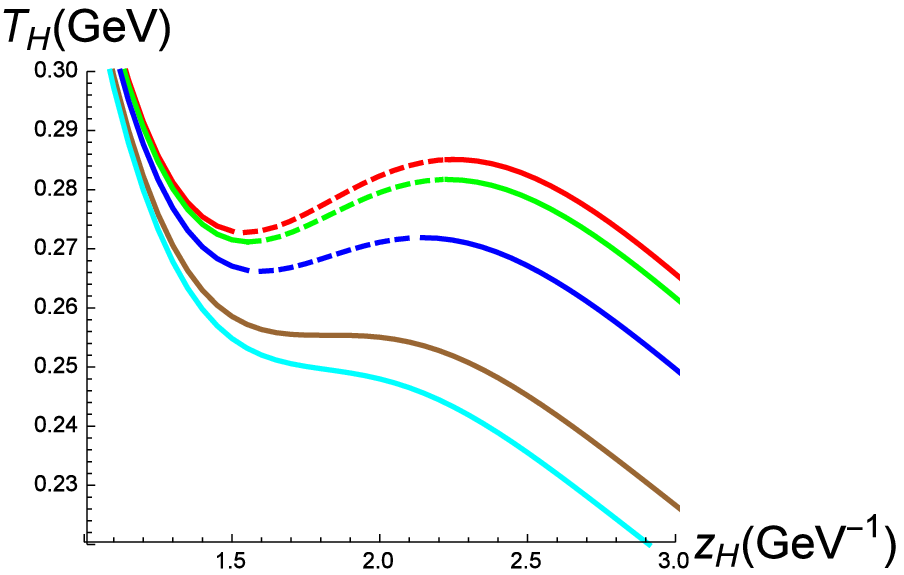}}
\caption{(color online) Hawing temperature $T_{H}$ as a function of the horizon radius $z_H$ for various chemical potential $\mu$. Solid lines denote stable branches and dashed ones denote unstable branches. {\bf Left}: For Case I, the red, green, blue, brown, cyan and magenta curves correspond to $\mu=0, 0.2, 0.4, 0.5, 0.6$ and $0.673 (\mu_c)$ respectively. {\bf Right}: For Case II, the red, green, blue, brown and cyan curves correspond to $\mu=0, 0.1, 0.2, 0.312 (\mu_c)$ and $0.35$ respectively. }
\end{figure}

In the left panel of Fig.~1, we plot the Hawking temperature $T_{\rm H}$ as a function of the horizon radius $z_H$ for various chemical potential $\mu$. From the figure, we can see that for small chemical potential $\mu<\mu_c = 0.673 {\rm GeV}$, there exist two branches of the black hole solution for a given temperature $T_{\rm H}$, one is stable (smaller $z_H$) while the other (larger $z_H$) is unstable. Beyond these two branches, there is also another solution, the thermal-AdS solution. There is a first-order phase transition between the stable black hole solution and the thermal-AdS, the so-called Hawking-Page transition.

By comparing their free energies, which are functions of the temperature $T_{\rm H}$ and the chemical potential $\mu$, we can obtain the transition temperature $T_{\rm HP}$ for various chemical potential, which is shown in the left panel of Fig.~2. At zero chemical potential, $T_{\rm HP} \simeq 0.264 {\rm GeV}$ which agrees well with the lattice result~\cite{Lucini:2003zr,Borsanyi:2012cr} (Actually, we use this condition to fix $\bar{a}$ to take the above value.). When the chemical potential exceeds $\mu_c$, the unstable branch of the black hole solution and the Hawking-Page transition disappear, thus defining a critical point $(\mu_c, T_c)$. From the phase diagram, we can see that the transition temperature decreases as the chemical potential increased, qualitatively agreeing well with the lattice results for heavy quarks.

By calculating the free energy of the $q\bar{q}$ pair holographically in this frame, one can observe that this Hawking-Page transition is dual to confinement-deconfinement transition of the boundary theory, with the thermal-AdS and AdS black hole correspond to confinement and deconfinement phases respectively.

\begin{figure}[!htbp]
\centering
\subfigure[$~~{\rm Case~I}$]{\includegraphics[width=0.45\textwidth]{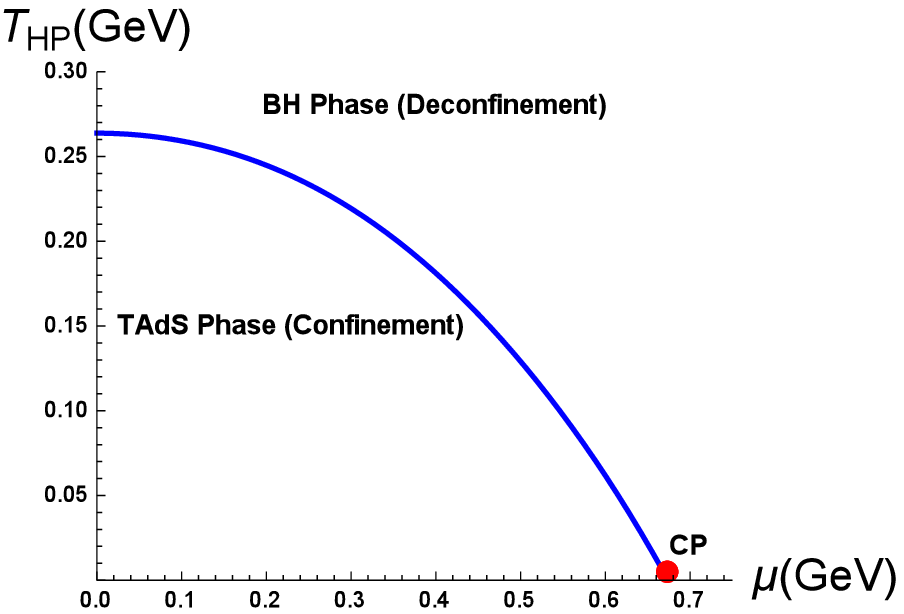}}\quad
\subfigure[$~~{\rm Case~II}$]{\includegraphics[width=0.45\textwidth]{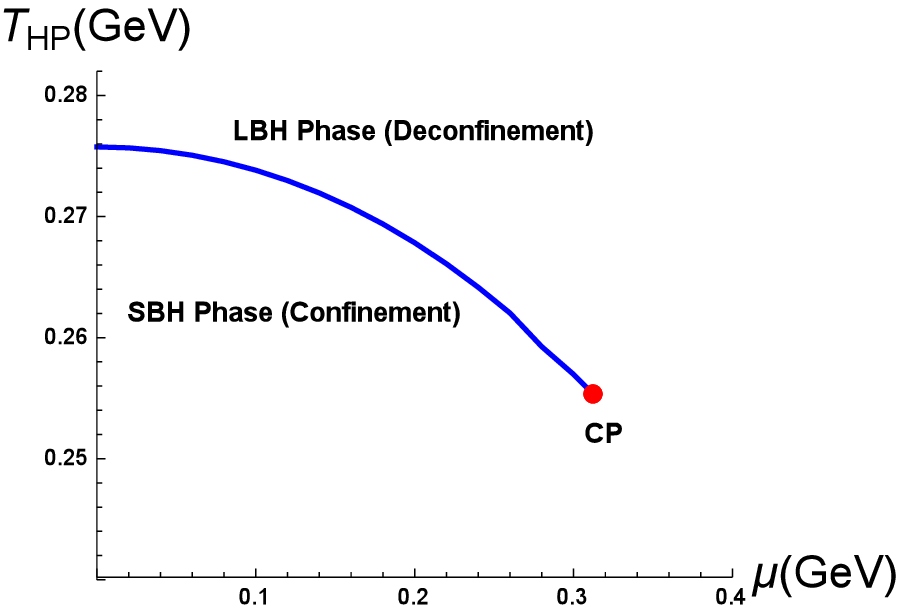}}
\caption{(color online) Phase transition temperature as a function of the chemical potential. {\bf Left}: At zero chemical potential, $T_{\rm HP} \simeq 0.264$. The critical point (cp) is at $(\mu_c, T_c)=(0.673,0.000159)$ with $T_c$ being very close to zero. {\bf Right}: At zero chemical potential, $T_{\rm HP} \simeq 0.275$. The cp is at $(\mu_c, T_c)=(0.312, 0.255)$.}
\end{figure}

\subsection{Case II: Specious-confinement/deconfinement phases}

In this case, the warped factor takes a more complicated form
\begin{eqnarray}
A(z) = -\frac{3}{4} \ln (a z^2 +1) + \frac{1}{2} \ln (b z^3+1) - \frac{3}{4} \ln (a z^4 +1),
\end{eqnarray}
with
\begin{eqnarray}
a = \frac{c}{9},\qquad b = \frac{5 c}{16}.
\end{eqnarray}

In the right panel of Fig.~1, the Hawking temperature of the AdS Black hole as a function of the horizon radius $z_H$ for various chemical potential is plotted. From the figure, we can see that, for a fixed temperature $T_H$, there at most exist three branches of solutions, two are stable (a large black hole (LBH) with small $z_H$ and a small black hole (SBH) with large $z_H$, denoted with solid lines) while the rest one is unstable (denoted with dashed line). There is also a similar Hawking-Page transition between the LBH and the SBH. Beyond these three branches, there also exists the thermal AdS solution. However, the thermal AdS solution always has a larger free energy than the dominated stable black hole solution for fixed temperature, so it will not involved into the phase transition.

By comparing the free energies of the LBH and the SBH, one can also obtain the transition temperature $T_{\rm HP}$ which depends on $\mu$. The full phase diagram is plotted in the right panel of Fig.~2. Note that there also exists a critical point $\mu_c$, above which the SBH branch and the Hawking-Page transition disappear.

It is suggested that the above large-small black hole transition is dual to the confinement-deconfinement transition in the dual boundary field theory. And the LBH and the SBH correspond to deconfinement and confinement phases respectively. However, as pointed out in Ref.~\cite{Dudal:2017max}, SBH is not a strict confinement phase as the corresponding expectation values of the Wilson loop and Polyakov loop do not precisely take the desired value.
However, in spite of this limitation, the corresponding thermodynamic properties in this phase, such as the entropy of the $q\bar{q}$ pair and speed of sound, agree well with the lattice QCD results in the confined phase. So, in Ref.~\cite{Dudal:2017max} the SBH phase is called specious-confinement phase, to distinguish it from the standard confined phase.

\section{Holographic subregion complexity}

In this section, we will study the subregion complexity using the subregion CV conjecture Eq.~(\ref{subCV}). For simplicity, we specify the subregion $B$ to be a strip:
\begin{eqnarray}
B: \quad x_1 \in \left[-\frac{l}{2}, \frac{l}{2} \right], \quad x_2 \in \left[-\frac{L_2}{2}, \frac{L_2}{2}\right], \quad x_3 \in \left[-\frac{L_3}{2}, \frac{L_3}{2}\right],
\end{eqnarray}
with $L_2, L_3 \gg l$ so we have translation symmetry along $x_2$ and $x_3$. Considering the symmetry, the HRT surface ${\gamma_B}$ can be parameterized as $z = z(x_1)$, and then the holographic entanglement entropy of $B$ is
\begin{eqnarray}\label{HEE}
S^{\rm HEE} = \frac{V_2}{2 G_5} \int_{0}^{l/2} dx_1 \frac{e^{3 A(z)}}{z^3} \sqrt{1+\frac{z'^2}{g(z)}},
\end{eqnarray}
where $V_2 \equiv L_2 L_3$ and the prime denotes derivative with respect to $x_1$. The HRT surface $\gamma_B$ is determined by extremizing $S^{\rm HEE}$ and thus satisfies the following equation
\begin{eqnarray}
2 g z z'' + \left(6 g -\frac{d g}{dz} z - 6 g \frac{d A}{d z} z \right) z'^2 - 6 g^2 \frac{d A}{dz} z + 6 g^2=0.
\end{eqnarray}
Generally, there are two kinds of extremal surfaces, one is connected and the other is disconnected. Which one is the real HRT surface with minimum area depends on $l$ (for details on holographic entanglement entropy, see Ref.~\cite{Dudal:2018ztm}).

The connected one has an $U$-shape and satisfies the following boundary conditions
\begin{eqnarray}
z(0)= z_\ast, \quad z'(0) =0,\quad z(\pm l/2) = 0,
\end{eqnarray}
where $z_\ast$ is the tip of $\gamma_B$ in the bulk. Noting that the Lagrangian in Eq.~(\ref{HEE}) does not depend on $x_1$ explicitly, so the associated Hamiltonian ${\cal H}$ is conserved thus leading to the following relation
\begin{eqnarray}
\frac{e^{3 A(z)}}{z^3 \sqrt{1+\frac{z'^2}{g(z)}}} = \frac{e^{3 A(z_\ast)}}{z_\ast^3}.
\end{eqnarray}
With this relation, we can express the strip length $l$ as a function of $z_\ast$
\begin{eqnarray}
l = 2\int_0^{z_\ast} dz \frac{z^3 e^{-3 A(z)}}{\sqrt{g(z) [z_\ast^6 e^{-6 A(z_\ast)} - z^6 e^{-6 A(z)}]}},
\end{eqnarray}
and the holographic entanglement entropy can be expressed as
\begin{eqnarray}
S^{\rm HEE}_{\rm con} = \frac{V_2}{2 G_5} \int_0^{z_\ast} dz \frac{z_\ast^3}{z^3} \frac{e^{3 A(z) - 3 A(z_\ast)}}{\sqrt{g(z) [z_\ast^6 e^{-6 A(z_\ast)} - z^6 e^{-6 A(z)}]}}.
\end{eqnarray}
And the corresponding subregion complexity can be written as
\begin{eqnarray}\label{Ccon}
{\cal C}_{\rm con} = \frac{V_2}{4 \pi G_5} \int_0^{l/2} dx_1 \int_0^{z(x_1)} dz \frac{e^{4 A(z)}}{z^4 \sqrt{g(z)}}.
\end{eqnarray}

In the disconnected case, the $U$-shaped extremal surface breaks into two pieces with fixed $x_1 = \pm l/2$ hanging from the boundary to the horizon or the center. Then we have
\begin{eqnarray}
S^{\rm HEE}_{\rm discon} &=& \frac{V_2}{2 G_5} \left[\int_0^{z_d} dz \frac{e^{3 A(z)}}{z^3 \sqrt{g(z)}} + \frac{e^{3 A(z_d)}}{2 z_d^3} l\right].\\
{\cal C}_{\rm discon} &=& \frac{l V_2}{8\pi G_5} \int_0^{z_d} dz \frac{e^{4 A(z)}}{z^4 \sqrt{g(z)}},\label{Cdiscon}
\end{eqnarray}
where $z_d=\infty$ or $z_d=z_h$, depending on whether the bulk geometry is thermal-AdS (TAdS) or AdS black hole (BH). From the above equations, we can see that for the TAdS phase, the entanglement entropy of the disconnected surface is independent of $\ell$ but the corresponding complexity is linear in $\ell$. While for the BH phase, the entanglement entropy of the disconnected surface and its corresponding complexity are both linear in $\ell$. In the following subsections, we will study the holographic complexity in the two cases mentioned in the last section. Generally, it is not possible to get analytical results, so we rely on numerical calculations. It should be noted that the integration in Eqs.~(\ref{Ccon})(\ref{Cdiscon}) are divergent, so we define renormalized complexity density as $\hat{\cal C} \equiv \frac{{\cal C} - {\cal C}_0}{V_2 \ell}$ where ${\cal C}_0$ is some reference value to cancel the divergence. For convenience, from now on we use the convention that $G_5 = 1$ and all physical quantities are in units {\rm GeV}.

\subsection{Case I: Standard confinement/deconfinement phases}

In the above discussions, we know that the holographic complexity depends on three parameters, the temperature $T_{\rm H}$, the chemical potential $\mu$ and the length scale $\ell$. Let us first discuss its behavior as varying $\ell$ in case I.

\begin{figure}[!htbp]
\centering
\includegraphics[width=0.5\textwidth]{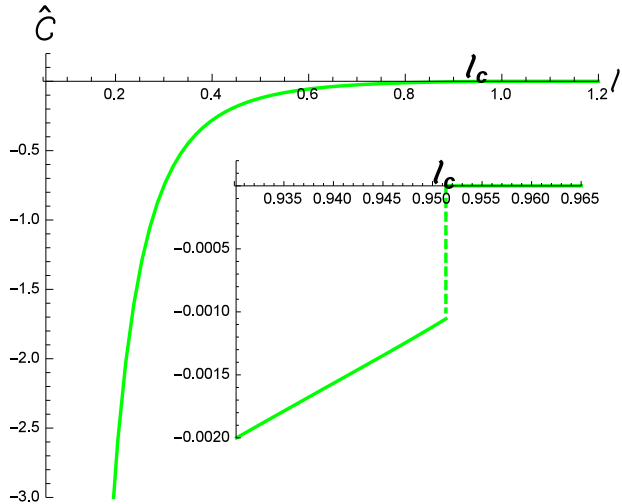}
%\subfigure[$~~{\rm Case~II}$]{\includegraphics[width=0.45\textwidth]{T-CaseII.eps}}
\caption{(color online) Renormalized complexity density $\hat{\cal C} \equiv \frac{{\cal C} - {\cal C}_0}{V_2 \ell}$ as a function of $\ell$ in the TAdS phase. ${\cal C}_0$ is the corresponding complexity of the disconnected extremal surface. Here the critical value $\ell_c \simeq 0.951$.}
\end{figure}

To calculate the holographic complexity in the two phases, TAdS phase and BH phase, we need first to find the HRT surface. As we stated above, there generally exists two kinds of competing extremal surfaces, one is connected and the other is disconnected. In the TAdS phase, for given temperature and chemical potential, which one is the real minimum surface-the HRT surface, depends on $\ell$. As observed in Ref.~\cite{Dudal:2018ztm}, there exists a critical value $\ell_c \simeq 0.951$ where a connected-disconnected transition happens. That is, when $\ell < \ell_c$, the HRT surface is the connected one; While $\ell > \ell_c$, it turns out to be the disconnected one.

In Fig.~3, the renormalized holographic complexity density $\hat{\cal C}$ as a function of the length scale $\ell$ is plotted. The reference value ${\cal C}_0$ is chosen as the corresponding complexity of the disconnected extremal surface. From the figure, we can see that $\hat{\cal C}$ experiences three stages as $\ell$ is increased. First, it grows quickly, then the growth rate drops sharply around $\ell \sim 0.4$ and after that $\hat{\cal C}$ approaches to zero slowly until at $\ell_c \simeq 0.951$ it jumps to zero. The discontinuous of $\hat{\cal C}$ at $\ell_c$ originates from the connected-disconnected transition.

Now let us turn to the BH phase. As observed in Ref.~\cite{Dudal:2018ztm}, for any $\ell$, the HRT surface is always the connected one. In Fig.~4, $\hat{\cal C}$ as a function of $\ell$ for various situations is plotted. Here the reference value ${\cal C}_0$ is the corresponding complexity of the disconnected extremal surface. From the figure, we can see that $\hat{\cal C}$ experiences two stages as one increases $\ell$, a quickly-growing stage at small $\ell$ and then a slowly-growing stage at large $\ell$. In the large scale limit $\ell \rightarrow \infty$, the subregion $B$ approaches the full system, and ${\cal C} = {\cal C}_0$ is nothing but the volume of the black hole exterior (up to an irrelevant factor). Different from that in the TAdS phase, $\hat{\cal C}$ is always continuous versus $\ell$ in the BH phase.

\begin{figure}[!htbp]
\centering
\subfigure[$~~\mu = 0$]{\includegraphics[width=0.45\textwidth]{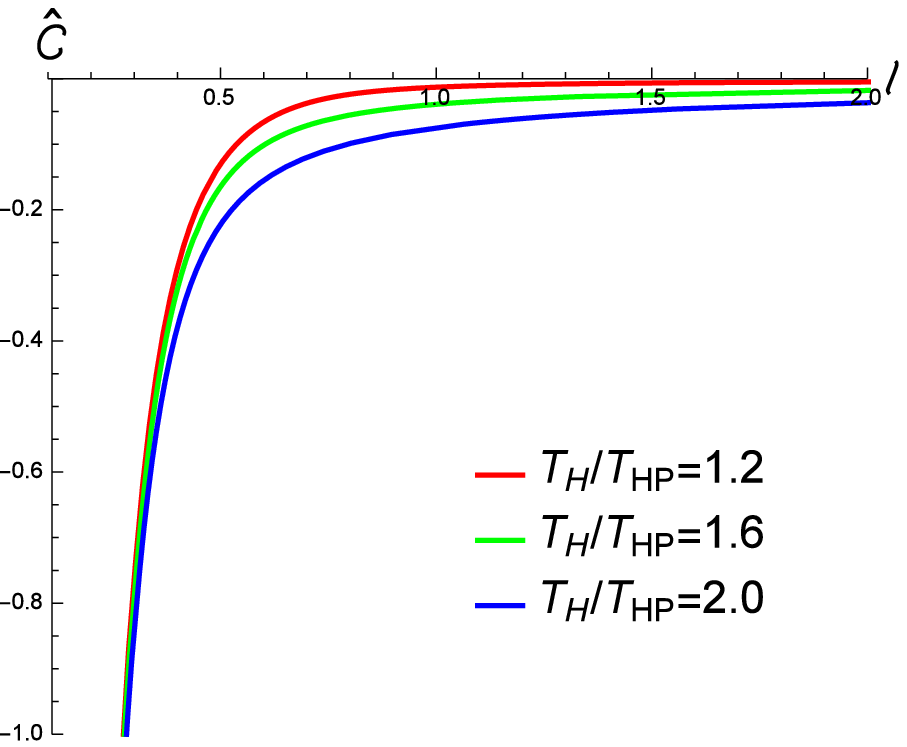}}\quad
\subfigure[$~~T_{\rm H}/T_{\rm HP} = 1.2$]{\includegraphics[width=0.45\textwidth]{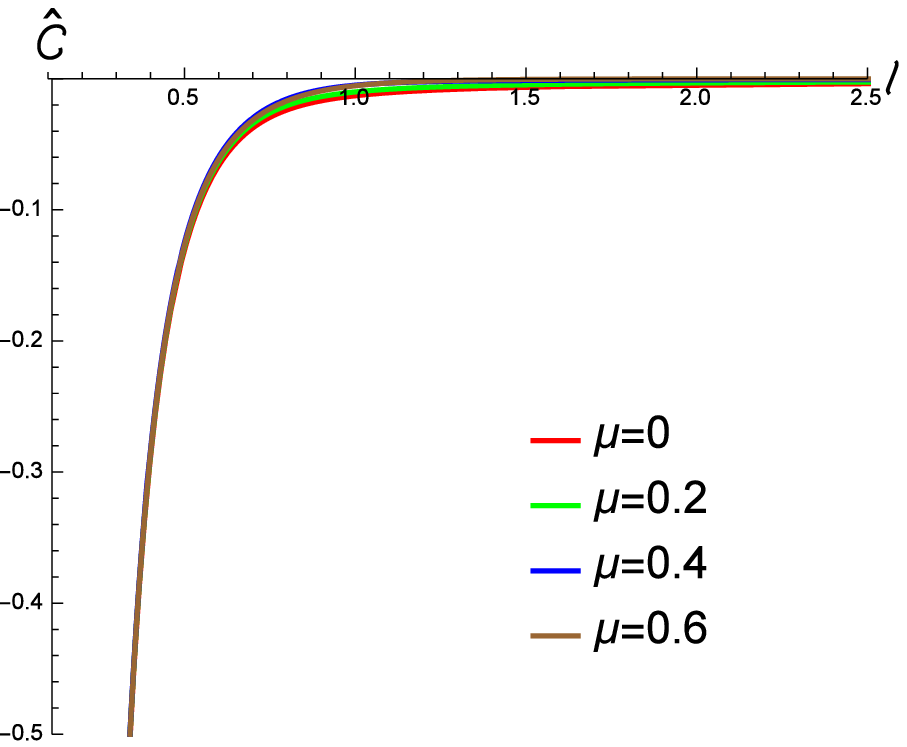}}
\caption{(color online)  Renormalized complexity density $\hat{\cal C} \equiv \frac{{\cal C} - {\cal C}_0}{V_2 \ell}$ as a function of $\ell$ in the BH phase. {\bf Left}: The chemical potential is fixed to be zero. {\bf Right}: The temperature is fixed to be $T_H/T_{\rm HP} =1.2$.}
\end{figure}

From the study in the last section, we see that there is a first-order phase transition between the TAdS phase and BH phase, which is interpreted holographically as a confinement-deconfinement transition of the dual boundary system. So, let us now move to study the behavior of the holographic complexity versus the temperature and the chemical potential to see if this phase transition leaves any imprint on the holographic complexity. The results are shown in Fig.~5. It should be noted that, for fixed chemical potential, there is a minimum temperature below which no BH solution exists. From this figure and Fig.~1, we can see that the holographic complexity is always discontinuous at the transition point. However, as the behaviors of holographic entanglement entropy~\cite{Dudal:2018ztm}, with only the holographic complexity density the transition point can not be pin-pointed exactly.

\begin{figure}[!htbp]
\centering
\includegraphics[width=0.5\textwidth]{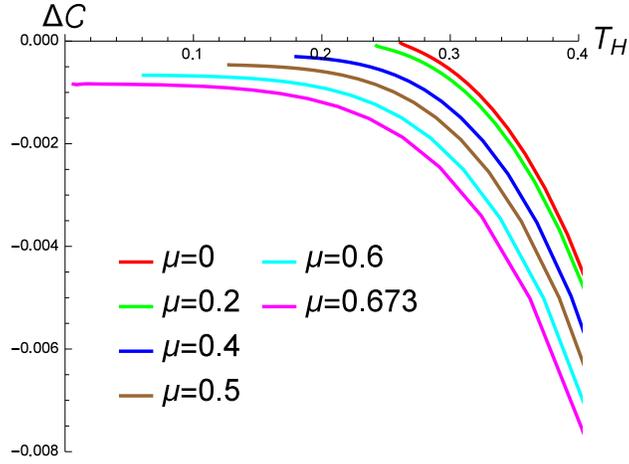}
\caption{(color online) The holographic complexity density difference between the TAdS phase and the BH phase $\Delta {\cal C} \equiv \frac{{\cal C}_{BH} - {\cal C}_{\rm TAdS}}{V_2 \ell}$ as a function of the temperature for various chemical potential with fixed $\ell =0.2$.}
\end{figure}

\subsection{Case II: Specious-confinement/deconfinement phases}

Let us now consider case II. In this case, there exists a first-order phase transition between a SBH and LBH for given chemical potential, and it is interpreted as a specious-confinement/deconfinement transition holographically.

\begin{figure}[!htbp]
\centering
\includegraphics[width=0.5\textwidth]{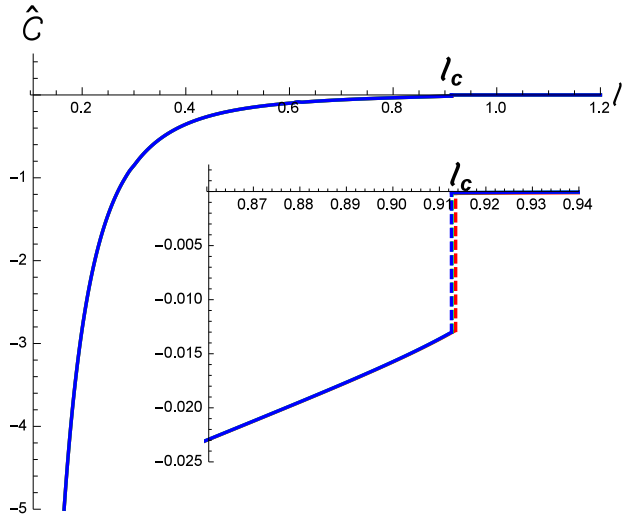}
\caption{(color online) The holographic complexity $\hat{\cal C} \equiv \frac{{\cal C} - {\cal C}_0}{V_2 \ell}$ as a function of $\ell$ in the SBH phase. The chemical potential is fixed $\mu=0$. Curves with red, green and blue colors correspond to $T_{\rm H}/T_{\rm HP} = 0.9, 0.8$ and $0.7$ respectively.}
\end{figure}

In the SBH phase, by studying the entanglement entropy at fixed temperature and chemical potential, it is found that the HRT surface is always connected~\cite{Dudal:2018ztm}. However, for a given length scale $\ell$, there are two competing branches of connected extremal surfaces, and which one is the HRT surface depends on $\ell$. There exists a critical value $\ell_c$ where the HRT surface sees a transition between the two branches.

In Fig.~6, the renormalized holographic complexity density as a function of $\ell$ in the SBH phase is plotted. The reference value ${\cal C}_0$ is chosen to be the corresponding complexity of the disconnected extremal surface. From the figure, we can see that $\hat{\cal C}$ exhibits a behavior similar to that in TAdS phase in case I. Again, $\hat{\cal C}$ is discontinuous at $\ell_c$ indicating the transition between the two connected branches.

\begin{figure}[!htbp]
\centering
\includegraphics[width=0.5\textwidth]{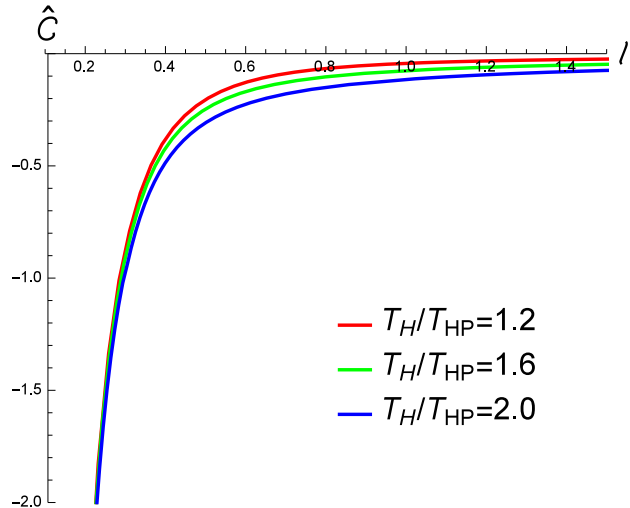}
\caption{(color online) The holographic complexity $\hat{\cal C} \equiv \frac{{\cal C} - {\cal C}_0}{V_2 \ell}$ as a function of $\ell$ in the LBH phase. The chemical potential is fixed $\mu=0$. Curves with red, green and blue colors corresponds to $T_{\rm H}/T_{\rm HP} = 1.2, 1.6$ and $2.0$ respectively.}
\end{figure}

Now let us turn to discuss the LBH phase. The renormalized holographic complexity density as a function of $\ell$ with fixed chemical potential is plotted in Fig.~7. The reference value ${\cal C}_0$ is again chosen as the corresponding complexity of the disconnected extremal surface. From the figure, we can see that $\hat{\cal C}$ exhibits a similar behavior as in the BH phase in case I. Again, $\hat{\cal C}$ is always continuous versus $\ell$.

\begin{figure}[!htbp]
\centering
\includegraphics[width=0.5\textwidth]{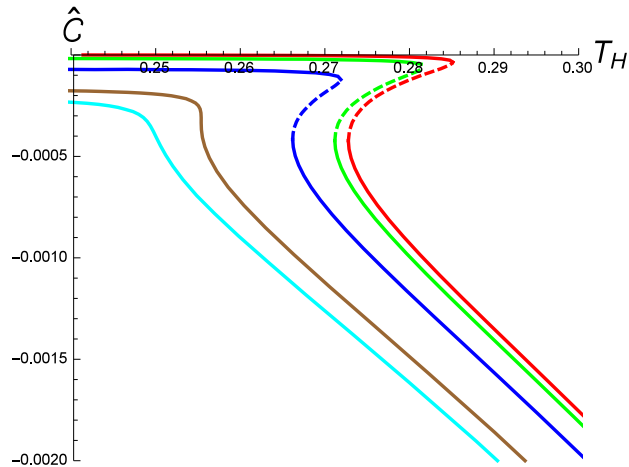}
\caption{(color online) The renormalized holographic complexity density $\hat{\cal C} \equiv \frac{{\cal C} - {\cal C}_0}{V_2 \ell}$ as a function of $T_{\rm H}$ with fixed $\ell =0.2$. Curves with red, green, blue, brown and cyan colors correspond to $\mu=0, 0.1, 0.2, 0.312 (\mu_c)$ and $0.35$ respectively. Solid lines denote stable branches while dashed lines denote unstable branches.}
\end{figure}

In Fig.~8, we show the behavior of $\hat{\cal C}$ as varying the temperature for fixed chemical potential to see if it can respect the Hawking-Page transition between the SBH and LBH phases. The reference value ${\cal C}_0$ is chosen to be the corresponding complexity of the disconnected extremal surface in TAdS background. From the figure and Fig.~1, we can see again that $\hat{\cal C}$ exhibits behavior characterizing the transition. That is, it is discontinuous at the transition point for $\mu<\mu_c$, and continuous for $\mu>\mu_c$ where the first-order SBH-LBH transition turns into a turnover. However, as in case I, the exact transition temperature can not be pin-pointed with only the help of the behavior of $\hat{\cal C}$.

\section{Summary and Discussions}

In this paper, we investigate the holographic complexity in a holographic QCD model that has been previously studied thoroughly in Refs.~\cite{Li:2011hp,Cai:2012xh,He:2013qq,Yang:2014bqa,Yang:2015aia,Dudal:2017max,Dudal:2018ztm}. By choosing appropriate forms of the gauge kinetic function $f(\phi)$ and warped factor $A(z)$ as well as values of parameters within, this model can reproduce the linear vector meson spectrum holographically which agrees well with the lattice results. Also, this model can realise well the confinement-deconfinement transition holographically. Two choices of the warped factor $A(z)$ are considered, case I and II. In case I, there is a first-order phase transition between TAdS and AdS BH. While in case II, there is a first-order phase transition between a SBH and a LBH. Both transitions can be interpreted as confinement-deconfinement transition of the dual system holographically.

By studying the behaviors of the renormalized holographic complexity density $\hat{\cal C}$ versus the length scale $\ell$, we find that in both cases, in the confinement phase (TAdS phase in case I and SBH phase in case II) there exists a critical value $\ell_c$ at which $\hat{\cal C}$ experiences a jump and thus is discontinuous. While in the deconfinement phase (BH phase in case I and LBH phase in case II), $\hat{\cal C}$ is always a continuous function of $\ell$. So, whether $\hat{\cal C}$ is discontinuous or continuous can be used to characterize the confinement or deconfinement phases. These behaviors are different from that of HEE. In Refs.~\cite{Dudal:2018ztm}, it is found that HEE $S^{\rm HEE}$ is always a continuous function of $\ell$ for both confined and deconfined phases. And it distinguishes the two phases by exhibiting different scaling behaviors versus $\ell$. In the confinement phase, one has~\footnote{It should be noted that this is not so strict for the SBH phase in case II.}
\begin{eqnarray}
\frac{\partial S^{\rm HEE}}{\partial \ell } \propto \left\{
\begin{array}{ll}
\frac{1}{G_5^0} = {\cal O} (N^0) & \quad {\rm for} \ \ell > \ell_c,\\
\frac{1}{G_5} = {\cal O} (N^2) & \quad {\rm for} \ \ell < \ell_c.
\end{array}
\right.
\end{eqnarray}
While in the deconfinement phase, one always has
\begin{eqnarray}
\frac{\partial S^{\rm HEE}}{\partial \ell } \propto
\frac{1}{G_5^0} = {\cal O} (N^0).
\end{eqnarray}

We also study the behaviors of $\hat{\cal C}$ versus the temperature and the chemical potential to see if it can respect the confinement-deconfinement transition. We find that $\hat{\cal C}$ always shows a behavior characterizing the transition. More precisely, it is always discontinuous at the transition temperature for $\mu<\mu_c$, and continuous for $\mu>\mu_c$ where first-order transition turns into a turnover. However, it should be noted that the transition temperature can not be pin-pointed with only the help of the behavior of $\hat{\cal C}$. These behaviors are similar to that of HEE~\cite{Dudal:2018ztm}.

All our results suggest that, as holographic entanglement entropy, the holographic complexity density can also be a good parameter to characterize the corresponding phase structures. In this paper, we only consider the subregion CV conjecture in the semi-analytical holographic QCD model. Whether our claim still holds for other situations, for example for the subregion CA conjecture or for other holographic QCD models, needs further investigations.

\section*{Acknowledgement}

We thank the hospitality of the Center for Gravitation and Cosmology of Yangzhou university during the visit where this project was initiated. This work was supported by National Natural Science Foundation of China (No. 11605155).

\end{document}